\documentclass[preprint,aps,prd,amsmath,amssymb]{revtex4}
\usepackage{graphicx}
\usepackage{color}
\usepackage[latin1]{inputenc}
\usepackage{ulem}
\newcommand{\be}{\begin{eqnarray}}
\newcommand{\beq}{\begin{equation}}
\newcommand{\eeq}{\end{equation}}
\newcommand{\ee}{\end{eqnarray}}

\newcommand{\bmp}{\noindent\begin{minipage}{16cm}}
\newcommand{\emp}{\end{minipage}\vskip 7mm} 

\newcommand{\pp}{{\bf p}}
\newcommand{\rr}{{\bf r}}
\usepackage{graphicx}
\usepackage{dcolumn}
\usepackage{bm}
\usepackage{amsmath}
\usepackage{amsfonts}
\usepackage{bbm}
\usepackage{subfigure}

\def\drawbox#1#2{\hrule height#2pt
        \hbox{\vrule width#2pt height#1pt \kern#1pt
              \vrule width#2pt}
              \hrule height#2pt}

\def\Asym#1#2{\vcenter{\vbox{\drawbox{#1}{#2}
              \kern-#2pt 
              \drawbox{#1}{#2}}}}

\begin{document}
\title{Zero Sound in Neutron Stars with Dense Quark Matter under Strong Magnetic Fields}
\author{Chris {\sc Kouvaris}}
\email{kouvaris@nbi.dk}
 \affiliation{The Niels Bohr Institute,
Blegdamsvej 17, DK-2100 Copenhagen \O, Denmark}


\begin{abstract}
We study a neutron star with a quark matter core under extremely strong magnetic fields.
We investigate the possibility of an Urca process as a mechanism for the cooling of such a star. We found that apart from very particular cases, the Urca process cannot occur. We also study the stability of zero sound modes under the same conditions. We derive limits for the coupling constant of an effective theory, in order the zero sound to be undamped. We show that zero sound modes can help kinematically to facilitate a cooling process.
\end{abstract}

\maketitle

\section{Introduction}
The cooling of neutron stars is a fascinating subject that is far from being well understood. There are several mechanisms that contribute to the cooling of a star. Generally, neutron stars cool down by emission of neutrinos from the whole volume, or photons from the surface. In particular, neutrons stars with nuclear matter cool down initially due to neutrino emission via the so-called modified Urca process, where the emissivity scales as temperature $T^8$, and later on, when the temperature becomes sufficiently small, due to photon emission from the surface of the star, that scales as $T^4$. In the case of nuclear matter, the direct Urca process is kinematically forbidden, because energy and momentum cannot be conserved simultaneously
~\cite{Baym:1975mf,Baym:1978jf,Pethick:1991mk,Yakovlev:2000jp,Yakovlev:2004iq,Prakash:2000jr,Page:2004fy,Yakovlev:2004yr,Shapiro:1983du}.

On the other hand, neutron stars with quark matter, can facilitate the direct Urca process. This was demonstrated by Iwamoto~\cite{Iwamoto:1980eb,Iwamoto2}. The emissivity of the direct Urca scales as $T^6$, which makes quark stars to cool faster than the corresponding ones with nuclear matter. The calculation of Iwamoto  assumes that quark matter is gapless. Direct Urca can also occur in dense nuclear matter with nonzero hyperon density~\cite{Glendenning:1984jr}, pion condensation~\cite{Bahcall:1965zzb}, and kaon condensation~\cite{Kaplan:1986yq}. Direct Urca can occur in all phases of quark matter, as long as the quarks are gapless. In sufficiently dense quark matter, quarks might pair in a nontrivial way, forming Cooper pairs. This is the phenomenon of Color Superconductivity~\cite{Alford:1997zt,Rapp:1997zu}. In this case, fermionic excitations might acquire an energy gap, thus making impossible the direct Urca process for phases like the Color-Flavor-Locking (CFL)~\cite{Alford:1998mk}. However in several cases, some fermionic excitations might remain gapless, and if kinematically allowed, they can facilitate a direct Urca process~\cite{Alford:2004zr}.

The case of neutron stars with strong magnetic fields is an interesting one. Although weak magnetic fields do not seem to influence a lot the cooling process, strong magnetic fields might alter the picture completely. This is because charged particles populate Landau levels. Under strong magnetic fields, where $eB\sim \mu^2$ ($\mu$ being the characteristic chemical potential), only the very first Landau levels are occupied. So far, several neutron stars have been observed with magnetic fields of $10^{14}$-$10^{15}$ Gauss. It is possible that the magnetic fields inside the star could be even higher. The emissivity of the direct Urca process for the case of a neutron star made of nuclear matter with strong magnetic fields was studied in~\cite{Leinson:1998yr,Baiko:1998jq}. It was shown that for $eB>\mu^2$, direct Urca reactions are open for an arbitrary proton concentration. Due to the fact that direct Urca has in general a larger emissivity compared to modified Urca, neutron stars with strong magnetic fields will cool much faster than stars without. By comparison of observational data, constraints can be imposed regarding the strength of the magnetic fields in the neutron stars, or vice versa, predictions can be made for anomalously cold neutron stars.

Although the case of neutron stars made of nuclear matter with strong magnetic fields has been studied and is
 somewhat understood, the case where the neutron star is made (partially) of quark matter
 under strong magnetic fields has not been studied to the same extend. Color superconducting quark matter under high magnetic fields has been studied in~\cite{Fukushima:2007fc}.
If quark matter exists inside neutron stars,
 in order to fill only the lowest Landau level, magnetic fields $qB\sim \mu^2$ are required. For a quark chemical potential of the order of 100 MeV, which is expected for quark matter in a neutron star, the corresponding magnetic field is of
 the order of $10^{18}$ Gauss. As we already mentioned, the highest magnetic fields so far observed in magnetars are $10^{14}-10^{15}$ Gauss. However, it is probable that stronger magnetic fields develop in their interior. From this point of view, the study of strong magnetic fields inside a neutron star with quark matter becomes an interesting problem.

 In this paper we shall argue that the direct Urca does not occur in general in this case. In fact we shall show that at extremely high magnetic fields, the ``triangle inequalities'', i.e. the simultaneous conservation of energy and momentum is not possible and therefore direct Urca is not open, with the exception of very specific choices of the chemical potentials.

This paper is organized as follows: In section II we show why direct Urca is not open in general for quark matter in strong magnetic fields. In addition, we point out under what specific conditions such a reaction can take place. In section III we study the zero sound modes of quark matter under strong magnetic fields and we show how they can help in satisfying the triangle inequalities, making possible a cooling reaction. We conclude in section IV.
\section{The Impossibility of Direct Urca}
The direct Urca reaction in the case of quark matter consists of two reactions
\beq u + e \rightarrow d + \nu, \label{a1}\eeq \beq d \rightarrow u + e + \bar{\nu}. \label{a2}\eeq
A similar reaction with a strange quark instead of a down, can also take place. In the presence of a magnetic field,
charged particles fill the Landau levels. The wavefunction of a charged particle is
\beq \psi =\frac{1}{(L_yL_z)^{1/2}}\exp[-iE_{\nu}t+ip_yy+ip_zz]u^{\uparrow \downarrow}, \eeq
where
 \begin{displaymath}
  u^{\uparrow}= \frac{1}{[2E_{\nu}(E_{\nu}+m)]^{1/2}}
  \left ( \begin{array}{c}
   (E_{\nu}+m)I_{\nu,p_y}(x)\\
  0 \\
  p_z I_{\nu,p_y}(x) \\
  -i(2 \nu qB)^{1/2} I_{\nu-1,p_y}(x)
  \end{array} \right ),
  \end{displaymath}
 and
 \begin{displaymath}
  u^{\downarrow}= \frac{1}{[2E_{\nu}(E_{\nu}+m)]^{1/2}}
  \left ( \begin{array}{c}
   0 \\
   (E_{\nu}+m)I_{\nu -1,p_y}(x)\\
  i(2 \nu qB)^{1/2} I_{\nu,p_y}(x) \\
  -p_z I_{\nu -1,p_y}(x)
  \end{array} \right ).
  \end{displaymath}
  The symbols $\uparrow$ and $\downarrow$ denote up and down spin states. We have chosen the magnetic field $B$ to be along the $z$ axis. The symbol $\nu$ indicates the Landau level, running from 0,1,2,.... $L_y$, and $L_z$ are nonphysical normalization lengths for the $y$ and $z$ axis, and they should drop out at the end of the day. The function $I_{\nu, p_y}$ is given by
  \beq I_{\nu,p_y}(x)=\left (\frac{qB}{\pi} \right )^{1/4} \frac{1}{(2^{\nu} \nu !)^{1/2}} \exp \left [ -\frac{1}{2}qB \left (
  x-\frac{p_y}{qB} \right )^2 \right ]H_{\nu} \left [ (qB)^{1/2} \left (x- \frac{p_y}{qB} \right ) \right ], \eeq
 where $H_{\nu}$ is the well known Hermite polynomial. We have assumed that the electric charge $q$ is positive, but similar formulas can be written down for a negative electric charge. The energy eigenvalues are $E_{\nu}=(p_z^2+m^2+2\nu qB)^{1/2}$, where $m$ is the mass of the particle. Here we are going to assume that $eB\sim \mu^2$ and therefore all the charged particles occupy the lowest Landau level, i.e. $\nu=0$. If the quark matter is in $\beta$ equilibrium,
 \beq \mu_d=\mu_u +\mu_e. \eeq In order for (\ref{a1}) to take place, momentum along the $z$ axis and energy should be conserved. This leads to
 \beq \sqrt{p_{zu}^2+m_u^2} + \sqrt{p_{ze}^2+m_e^2}=\sqrt{p_{zd}^2+m_d^2} +E_{\nu} \quad \text{for energy}, \label{b1} \eeq
 \beq p_{zu}+p_{ze}=p_{zd}+p_{\nu} \quad \text{for $z$ momentum}. \label{b2} \eeq
Similar equations hold for the case of the strange quark, where we can substitute $d\rightarrow s$ in Eqs.~(\ref{a1},~\ref{a2},~\ref{b1},~\ref{b2}). Obviously, the $\beta$ reactions with the strange quark are somewhat suppressed due to the Cabibbo angle. Since we have assumed that all the charged particles occupy only the lowest Landau level, this means that the particles fill all the states between $-p_F$ to $p_F$, where $p_F=\sqrt{\mu^2-m^2}$. Having chosen the magnetic field along the $z$ axis, up quarks in the lowest Landau level have their spin aligned along the positive $z$ axis, whereas down and strange quarks, as well as electrons have their corresponding spins along the $-z$. The electroweak sector involves only left handed particles, and therefore at the limit where the mass of the particles goes to zero, only the up quarks with momentum $-p_{Fu}$, down quarks with momentum $p_{Fd}$, and electrons with $p_{Fe}$ would participate in $\beta$ reactions. However, since the particles do have a mass, up quarks with $+p_{Fu}$, as well as down quarks with $-p_{Fd}$ and electrons with $-p_{Fe}$ can participate, although the amplitude will be suppressed as $m/\mu$, with $m$ and $\mu$ being the corresponding masses and chemical potentials of the particles. Similarly for the reaction~(\ref{a2}) we have the following equations from energy and momentum conservation
\beq \sqrt{p_{zu}^2+m_u^2} + \sqrt{p_{ze}^2+m_e^2}+E_{\bar{\nu}}=\sqrt{p_{zd}^2+m_d^2}  \quad \text{for energy}, \label{c1} \eeq
 \beq p_{zu}+p_{ze}+ p_{\bar{\nu}}=p_{zd} \quad \text{for $z$ momentum}. \label{c2} \eeq

Now we shall argue that the set of Eqs.~(\ref{b1},~\ref{b2},~\ref{c1},~\ref{c2}) cannot be satisfied in general. Firstly, let's consider the case where up, down and electrons have velocities aligned along the $+z$ axis. Quark matter if present inside neutron stars, is expected to be at high densities, which means that $m<<\mu$ and $m<<p_F$. We can expand Eqs.~(\ref{b1},~{\ref{c1}) around the corresponding Fermi momenta and we get respectively
\beq p_{Fu} +\frac{m_u^2}{2\mu_u}+p_{Fe}+\frac{m_e^2}{2\mu_e}=p_{Fd}+\frac{m_d^2}{2\mu_d}+E_{\nu}, \label{d1} \eeq \beq p_{Fu} +\frac{m_u^2}{2\mu_u}+p_{Fe}+\frac{m_e^2}{2\mu_e}+E_{\bar{\nu}}=p_{Fd}+\frac{m_d^2}{2\mu_d}. \label{d2} \eeq
Subtracting (\ref{b2}) from (\ref{d1}) and (\ref{c2}) from (\ref{d2}) ($|p_z|=p_F$) , we get
\beq E_{\nu}-p_{\nu}=-A, \label{e1}\eeq and \beq E_{\bar{\nu}}-p_{\bar{\nu}}=A, \label{e2}\eeq where $$A=\frac{m_d^2}{2\mu_d}-\frac{m_u^2}{2\mu_u}-\frac{m_e^2}{2\mu_e}.$$ We can easily see now that the Urca process cannot occur in this case. First of all, since the energy of neutrinos is $\sim T$, the above equations require also $A\sim T$, which is not true in general. However even if this is true, since $E_{\nu,\bar{\nu}}\ge p_{\nu,\bar{\nu}}$, Eqs.~(\ref{e1}), and~(\ref{e2}) cannot be satisfied simultaneously provided $A\neq 0$. Now we can consider a case where one of the particles has velocity with opposite sign with respect to the rest. The most important case is to consider the up quark with $p_{zu}=-p_{Fu}$. In this case there is no $m/\mu$ suppression of the amplitude. Because of the flip of the sign of the momentum, Eqs.~(\ref{e1}, \ref{e2}) become
\beq E_{\nu}-p_{\nu}+A=2p_{Fu}, \eeq
\beq E_{\bar{\nu}}-p_{\bar{\nu}}+2p_{Fu}=A. \eeq This implies that the reactions (\ref{a1},~\ref{a2}) can take place if $2p_{Fu}=A$. This relation can be rewritten if we expand around $\mu_u$ as
\beq 2\mu_u= \frac{m_d^2}{2 \mu_d}+\frac{m_u^2}{2 \mu_u}-\frac{m_e^2}{2 \mu_e}. \label{f16}\eeq
Although in principle such a relation is not forbidden, it is highly unlike to be satisfied, as it requires very low chemical potential for the up quark. This can be improved by considering strange quarks instead of down, because the mass of the strange quark is much larger compared to the down. In that case, \beq \mu_u \simeq \frac{m_s^2}{4\mu_s}, \label{f18} \eeq where we omitted the subleading terms of (\ref{f16}). If it is the electron and not the up quark that flips velocity, we can carry the same analysis, concluding $\mu_e=m_s^2/(4\mu_s)$. However, in this case the reaction will be suppressed by factors of $(m_u m_e)^2/(\mu_u \mu_e)^2$. Eq.~(\ref{f18}) (or the corresponding with $e$ instead of $u$) is very restrictive and hard to be satisfied. It means that only for a very specific value of the chemical potential, the Urca process can occur.

We can now impose overall electric neutrality for the quark matter. This implies that $(2/3)n_u=(1/3)n_d+(1/3)n_s +n_e$, where $n$ stands for the number density of the particular particle. To calculate the number density we have to make the substitution $$\int \frac{dp_xdp_y}{4\pi^2}\rightarrow \frac{qB}{2\pi}.$$ Upon making this substitution, the number density is \beq n=\frac{qB}{2\pi^2}p_F. \label{numberdensity}\eeq After expanding for small masses, the neutrality condition becomes
\beq  \mu_s+ \mu_d +9 \mu_e-4\mu_u=\frac{m_s^2}{2\mu_s}+\frac{m_d^2}{2\mu_d}+9\frac{m_e^2}{2\mu_e}-\frac{2m_u^2}{\mu_u}. \label{elne} \eeq If we assume that the reactions of (\ref{a1}), and (\ref{a2}) take place for the strange quark, thermal equilibrium suggests $\mu_s=\mu_u+\mu_e$. As we showed, the same reactions cannot hold simultaneously for the down quarks, because it would imply a chemical potential for the quarks of the order of $m_d$ or less. Therefore $\beta$ equilibrium between up and down quarks cannot be established through Eqs.~(\ref{a1},~\ref{a2}). In this case, we have four parameters $\mu_u$, $\mu_d$, $\mu_s$, and $\mu_e$, where one is fixed by Eq.~(\ref{f18}), one by chemical equilibrium between strange, up and electrons, and one by electric neutrality. There is one left free parameter, and in order for direct Urca to occur, the neutrality condition (after using the chemical equilibrium condition, Eq.~(\ref{f18}), and having kept only the first term from the right hand side of Eq.~(\ref{elne})), reads $10\mu_s +\mu_d=15m_s^2/(4\mu_s)$. If for any reason the down quarks are in equilibrium with the other species, $\mu_s=\mu_d$. In this case, the above condition becomes $\mu_s=m_s\sqrt{15/44}$.
This is of no interest, because direct Urca occurs only for a single value of $\mu_s$, which is in addition unrealistically low. Furthermore, in this case the expansion we have done is not valid any more since $\mu_s$ is smaller than $m_s$. If the triangle inequalities are satisfied for the electron instead for the up quark, i.e. $\mu_e=m_s^2/(4\mu_s)$, the condition for direct Urca is $\mu_d=3\mu_s- 11m_s^2/(4\mu_s)$. Again, in case where $\mu_s=\mu_d$, the condition reads $\mu_s=m_s\sqrt{11/8}$. As we mentioned, it is not expected that $\mu_s=\mu_d$, since chemical equilibrium between up, down and electrons cannot be established in general. Our conclusion is that the direct Urca process can occur only if Eq.~(\ref{f18}) is satisfied (with either the up quark or the electron) and $\mu_d$ is such that satisfies the neutrality condition.

\section{Zero Sound Modes}
As we argued, in the case of quark matter at extremely strong magnetic fields, the direct Urca process cannot occur in general. Only under very specific conditions and probably unrealistic, such a process can take place. In the case of nuclear matter (without magnetic fields), the direct Urca does not occur either (unless there is nonzero hyperon density or pion, or kaon condensation), because the triangle inequalities are not satisfied. Instead, the modified Urca, with the presence of a bystander particle takes place. In our consideration, a bystander particle cannot help. This is because even the extra particle, either it is up, down, strange or electron, it should also be aligned along the $z$-axis, and therefore it is difficult again to satisfy the triangle inequalities. As in the direct Urca, perhaps such a process can occur, but again only under very specific values of the chemical potentials.

In this section we are going to study the zero sound in this particular setup, and we are going to investigate if such modes can help establish a cooling process by ensuring that the triangle inequalities are satisfied. First let's consider how the zero sound modes, if undamped, can facilitate simultaneous conservation of energy and momentum. Let's assume for the moment that the zero sound modes are undamped and that their velocity is $v_0<1$.
We can consider the following reactions (which are modifications of the $\beta$ reactions~(\ref{a1},~\ref{a2}) with the inclusion of a zero sound)
\beq u + e \rightarrow d + \nu + \text{zs}, \label{ma1}\eeq \beq d +\text{zs} \rightarrow u + e + \bar{\nu}, \label{ma2}\eeq where zs stands for zero sound. The equations that correspond to the previous Eqs.~(\ref{e1},~\ref{e2}) are respectively
\beq E_{\nu}-p_{\nu}-(1-v_0)k=-A, \label{me1}\eeq and \beq E_{\bar{\nu}}-p_{\bar{\nu}}=A-(1-v_0)k. \label{me2}\eeq
$k$ is the momentum of the zero sound, and the term $-(1-v_0)k$ is $\omega-k=v_0k-k=-(1-v_0)k$, where we used the dispersion $\omega=v_0k$ for the zero sound. From Eqs.~(\ref{me1},~\ref{me2}), we see that energy and momentum can be conserved, if we require $(1-v_0)k=A$. In some sense, the above mechanism corresponds to a ``modified Urca'' with the bystander being the zero sound, rather than a particle. We have already mentioned that modified Urca with a charged particle as a bystander would not help, as it would face the same limited phase space with direct Urca due to the fact that the bystander will have to be aligned along the $z$ axis with a Fermi momentum. Here, we see that in principle, zero sound modes can make this ``modified'' version of the Urca process to occur at high magnetic fields. Of course for such a process to take place, the zero sound modes should be undamped. In the rest of this section we are going to study the zero sound, in the case of quark matter under strong magnetic fields, using an effective Lagrangian.

The zero sound is a collective mode in Fermi liquids, involving particles near the Fermi surface. It was predicted by Landau~\cite{Landau} in 1959 and it was observed experimentally in liquid $He^3$. The relativistic extension of the Landau Fermi liquid theory (applicable for example at high density QCD) was developed by Baym and Chin~\cite{Baym:1975va}. In that paper, the zero sound velocity was calculated for a relativistic Fermi liquid using different effective models. Here, we calculate the zero sound velocity for a relativistic Fermi liquid that is under ultra strong magnetic fields, and therefore all charged particles fill only the lowest Landau level.

The Boltzmann equation that governs the number density in the collisionless case is~\cite{Landau2}
  \beq \frac{\partial n}{\partial t}+ \frac{\partial \epsilon}{\partial \pp}\frac{\partial n}{\partial \rr}-
  \frac{\partial \epsilon}{\partial \rr}\frac{\partial n}{\partial \pp}=0, \label{boltz}\eeq where $\epsilon$ and $n$ are the energy and density distribution of the quasiparticle. We consider small deviations from the equilibrium distribution
  \beq \hat{n}(\pp,\rr,t)=n_0(\pp)+\delta\hat{n}(\pp,\rr,t). \eeq The infinitesimal change in the energy is
  \beq \delta \hat{\epsilon}=\text{tr}'\int \hat{f}\delta \hat{n}'d\tau', \label{de} \eeq where $\int d \tau'$ is the integral over the phase space, the trace is over spin states, and $\hat{f}$ is the usual Landau interaction function. If we expand Eq.~(\ref{boltz}) and keep only up to first order variations, we have
  \beq \frac{\partial \delta \hat{n}}{\partial t}+\frac{\partial \epsilon_0}{\partial \pp}.\frac{\partial \delta \hat{n}}{\partial \rr}-\frac{\partial \delta \hat{\epsilon}}{\partial \rr}.\frac{\partial n_0}{\partial \pp}=0, \label{eq}\eeq with $\epsilon_0$ and $n_0$ being the energy and number density equilibrium distributions. Because all the charged particles are in the lowest Landau level, the equilibrium distributions for the density and the energy depend only on the $z$ component. $\partial \epsilon_0/\partial \pp={\bf v}$, where ${\bf v}=v_F{\bf n}$, ${\bf n}$ being the unit vector in the direction of $\pp$. It is understood that ${\bf n}$ can be  along the $\pm z$ axis. On the other hand, $\partial n_0/\partial \pp=-{\bf v}\delta(\epsilon-\epsilon_F)$. We are seeking wave solutions of the form
  \beq \delta \hat{n}=\delta(\epsilon-\epsilon_F){\bf \hat{\nu}(n)}e^{i({\bf k.r}-\omega t)}. \label{sol}\eeq

  Before we investigate the solution of Eq.~(\ref{eq}), we derive the Fermi interaction function $\hat{f}$, as it is quite different from the case where the magnetic field is absent. Let's assume that the Lagrangian includes an interaction term of the form
  \beq L_{\text{int}}=ig\bar{\psi}\gamma^{\mu}\psi A_{\mu}, \eeq where $A$ is a massive vector field. In this case we can write the amplitude of two fermions scattering by exchanging $A$ in the space coordinates as
  \beq T_{fi}=ig^2\int d^4x d^4x'j^{\mu}_{fi}(x)D(x-x')j_{\mu fi}(x'), \label{tfi} \eeq where $j^{\mu}_{fi}= \bar{\psi}\gamma^{\mu}\psi$ and the vector propagator $D$ is
  \beq D(x-x')=-\int \frac{d^4q}{(2\pi)^4}\frac{\exp[-iq.(x-x')]}{q^2-M^2}, \eeq where $M$ is the mass of the vector boson. We can focus on the case of a positive fermion (for example the up quark), where the wavefunction is given by Eq.~(3), with $\nu=0$ for the spinor, since we consider only the lowest Landau level. Let's denote by the subscripts 1 and 2 the incoming particles and 3, and 4 the outgoing ones. Eq.~(\ref{tfi}) involves integration over $d^4x d^4x' d^4q$. Because the spinor of the wavefunction in the lowest Landau level depends only on $x$, we can easily perform the integration over $dt'dy'dz'$ that gives the product of delta functions $(2\pi)^3L_yL_z\delta(q_0-E_2+E_4) \delta(q_y-p_{y2}+p_{y4}) \delta(q_z-p_{z2}+p_{z4})$. Integration over $dq_0dq_ydq_z$ is now trivial due to the delta functions, and finally after integrating over $dtdydz$, we get
  \beq T_{fi}=-i(2 \pi)^3g^2 \delta(E)\delta(p_y)\delta(p_z)\int dxdx'\frac{dq_x}{2\pi}\frac{e^{iq_x(x-x')}}{q^2-M^2}B_{\mu}(x)B^{\mu}(x'), \eeq
  where $B^{\mu}(x)= \bar{u}^{\uparrow}(x) \gamma^{\mu} u^{\uparrow}(x),$ and the delta functions denote overall energy and momentum ($y$ and $z$ axis) conservation. The Landau interaction function is defined as
  $T_{fi}=i(2\pi)^3 \delta(E)\delta(p_y)\delta(p_z)\hat{f}$. There are two distinct channels, we should consider. One is the direct forward scattering where $E_1=E_3$ and $E_2=E_4$ (with similar relations for the corresponding momenta), and the second is when $E_1=E_4$ and $E_2=E_3$ that corresponds to an energy and momentum exchange between the two particles. For the first case the integral \beq \int \frac{dq_x}{2\pi}\frac{1}{q^2-M^2}\exp[iq_x(x-x')]=-\frac{\exp[-M|x-x'|]}{2M}. \label{integ} \eeq The Landau interaction function is
  \beq \hat{f}(p_y,p_z;p_{y'},p_{z'})=\frac{g^2qB}{2M\pi} \int dxdx'e^{-M|x-x'|}e^{-qB(x-p_y/(qB))^2}e^{-qB(x'-p_{y'}/(qB))^2} \left (1-\frac{p_zp_{z'}}{EE'} \right ). \eeq
For convenience prime energy and momentum correspond to the ones with subscript 2 (in the direct amplitude subscript 4 as well).
We proceed now to calculate $\partial \delta \hat{\epsilon}/\partial \rr$ as we need it for Eq.~(\ref{eq}). The spin trace of Eq.~(\ref{de}) is trivial as all particles have spin up. Plugging in the solution of Eq.~(\ref{sol}), we have
\beq \frac{\partial \delta \hat{\epsilon}}{\partial \rr}=\int \frac{dp_{y'}}{2\pi}\frac{dp_{z'}}{2\pi}\hat{f}(p_y,p_z;p_{y'},p_{z'})\nu({\bf n'})i {\bf k}e^{i({\bf k.r}-\omega t)} \delta(E'-E_F). \eeq After integrating over $dp_{y'}dxdx'$ we get
\beq \frac{\partial \delta \hat{\epsilon}}{\partial \rr}=i {\bf k} e^{i({\bf k.r}-\omega t)}\frac{qB}{4\pi^2}\frac{g^2}{M^2}\int dp_{z'} \delta (E'-E_F)\nu({\bf n'}) \left (1-\frac{p_zp_{z'}}{EE'} \right ). \eeq There are only two directions for ${\bf n'}$, either up or down. After we perform the integration,
\beq \frac{\partial \delta \hat{\epsilon}}{\partial \rr}=i {\bf k} e^{i({\bf k.r}-\omega t)}\frac{qB}{4\pi^2}\frac{g^2}{M^2} \left (\hat{\nu}(p_z)\frac{m^2}{E_Fp_F}+\hat{\nu}(-p_z)\frac{2E_F^2-m^2}{E_Fp_F} \right ). \eeq We can now rewrite Eq.~(\ref{eq}) using the solution~(\ref{sol}) as
\beq (\omega - {\bf v.k}) \hat{\nu}(n)={\bf v.k}\frac{qB}{4\pi^2}\frac{g^2}{M^2} \left ( \hat{\nu}(n)\frac{m^2}{E_Fp_F}+\hat{\nu}(-n)\frac{2E_F^2-m^2}{E_Fp_F} \right ). \eeq This equation leads to a homogeneous system of two equations
\beq [\omega -v_Fk(1+x)]\hat{\nu}(+)-v_fkx\delta \hat{\nu}(-)=0, \eeq
\beq [\omega +v_Fk(1+x)]\hat{\nu}(-)+v_fkx\delta \hat{\nu}(+)=0, \label{dispersion2} \eeq
where $$x=\frac{qB}{4\pi^2}\frac{g^2}{M^2}\frac{m^2}{E_Fp_F},$$ $\delta=(2E_F^2-m^2)/m^2$, and $\pm$ denotes the direction of the momentum $p$~along or opposite to the direction of {\bf k}. The latter has two distinct directions along the $\pm~z$ axis. The determinant of this system of equations should be zero if a nontrivial solution exists. This gives the following dispersion relation
\beq \omega^2=v_F^2k^2[1+(1-\delta^2)x^2+2x]. \label{omega} \eeq Generally, in order for the zero sound to be undamped, its velocity should satisfy $v_0>v_F$. This condition can be easily proven upon making the requirement that the frequency $\omega$ has no imaginary part. However, in the case of a strong magnetic field, this condition is different. For $\omega$ to be real, Eq.~(\ref{omega}) gives
\beq x<\frac{1}{\delta-1}. \eeq
Assuming $m<<E_F$, the above equation becomes
\beq \frac{g}{2\pi}<\frac{M}{\sqrt{2qB}}.\eeq
 Unless the coupling constant satisfies the above equation, the zero sound modes will be damped. We should note that if we were to impose the general condition $v_0>v_F$, the upper bound for the coupling $g$ would have been $$ \frac{g}{2\pi}=\frac{M}{\sqrt{2qB}} \frac{m}{E_F},$$ which is suppressed compared to the one we derived above, by an extra factor of $m/E_F$. There are two points that we would like to stress here. The first one is the difference between our constraint about the undamped zero sound and the Landau general one ($v_0>v_F$) that holds in the case without magnetic field. The Landau constraint from the mathematical point of view comes from the requirement that $\omega$ has no imaginary part, which reflects the simple physical fact that zero sound modes with velocity less than $v_F$ lie in the continuum, and they are unstable as they can decay back to a particle  and a hole. With $v_0>v_F$, such a situation is avoided as the zero sound mode cannot decay to a particle-hole pair preserving energy and momentum simultaneously. The case we study here is qualitatively different. Because the particles fill only the lowest Landau level, the problem is one dimensional rather than three. The argument we used above, cannot be used now: there can be zero sound modes with velocity smaller than $v_F$ that cannot decay to particle-hole pair, because there is no continuum in this case. In the three dimensional case $\omega=v_F k \cos\theta$, with $\theta$ being the angle between the ${\bf v_F}$ and ${\bf k}$ vectors. For a proper choice of $\theta$, $\omega$ can take all the values between 0 and $\omega=|{\bf v_F}||{\bf k}|$. In our case though, because the problem is one dimensional, fermionic excitations above the Fermi surface can lie on the line $\omega=v_F k$, but not below (or above). Therefore zero sound  modes with $v_0<v_F$, cannot decay to a particle-hole pair. Our constraint of Eq. (42) or (43), show that there is a dynamical instability if it is not fulfilled, but it is not related to Landau damping.

The second point has to do with the notion of zero sound in one-dimensional systems. Generally, zero sound modes are viewed as distortions of the Fermi surface along the forward scattering direction, while first sound modes correspond to simple radial oscillations of the Fermi surface. It is obvious that such a distinction is not valid in the case we investigate, as the Fermi sphere is transformed to an one-dimensional column. From this point of view, the distinction between zero and first sound modes becomes fuzzy. However, we can still talk of zero or first sound with respect to the quantity $\omega \tau$, that is the energy of the mode times the mean time between collisions for the quasi-particle. $\omega \tau>>1$, meaning that the frequency of the mode is much larger than the frequency of collisions, corresponds to zero sound, whereas $\omega \tau<<1$ correspond to the case where the hydrodynamic limit is valid and the modes correspond to first sound. Generally, zero sound exists in Fermi liquids at low temperatures since $\tau \sim T^{-2}$. However in our case, because of the difference in the phase space, we should examine how $\tau$ scales. This is easy to do if we recall that the particles that can interact with each other are the ones that have momenta within $T$ from the Fermi surface. This means that the effective density of interacting particles is $\Delta n=qBT/\pi^2$. The effective cross section for a collision is $\sigma \sim (qB)^{-1}$, and the mean free path (and consequently $\tau$) should scale as $l\sim T^{-1}$. For energy $\omega$ larger than a few keV (which is a typical neutron star temperature), it is safe to say that $\omega \tau>>1$ and therefore we indeed have a zero sound mode.

We should also stress here, that the quasi-particle energy is given by
\beq \epsilon_p =\epsilon_p^0 + \int \hat{f}(p_y,p_z;p_{y'},p_{z'})n(p_{z'}) \frac{dp_{y'}}{2\pi}  \frac{dp_{z'}}{2\pi}. \eeq Plugging the expression of $\hat{f}$ for the direct amplitude and keeping in mind that
to lowest order $-p_F<p_{z'}<p_F$, we get
\beq \epsilon_p=\epsilon_p^0 +\frac{g^2}{M^2}n, \eeq where $n$ is defined in Eq.~(\ref{numberdensity}). This agrees with the
corresponding relativistic Fermi liquid (and zero magnetic field) formula of~\cite{Baym:1975va}~and with the result of~\cite{Ghosh:2004sn}. We should also mention that the zero sound velocity $v_0$ should be obviously smaller than the speed of light. It turns out that if $\delta^2(1-v_F^2)>1$, $v_0$ is smaller than one for any value of $x$ or correspondingly of the coupling constant $g$. If we choose a typical chemical potential of 500 MeV for a quark of mass 10 MeV (say down for example) $\delta^2(1-v_F^2)\sim 10^4>1$ which means that $v_0<1$ no matter what the coupling constant $g$ is.

  \begin{figure}[!tbp]
  \begin{center}

      \subfigure{\resizebox{!}{4.8cm}{\includegraphics{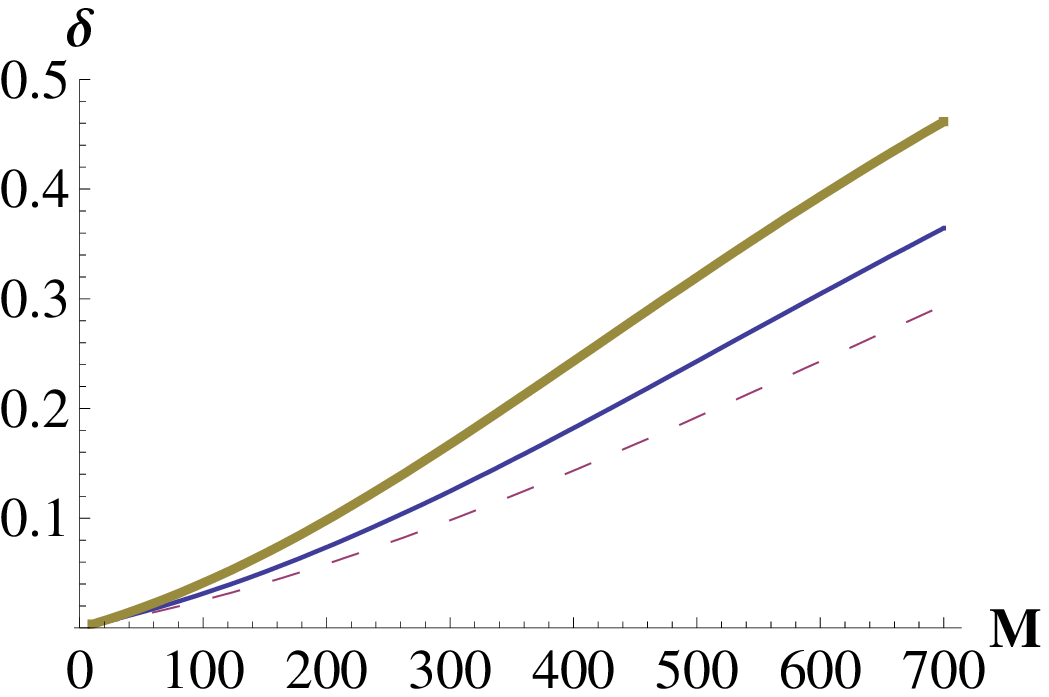}}} \quad
      \subfigure{\resizebox{!}{4.8cm}{\includegraphics{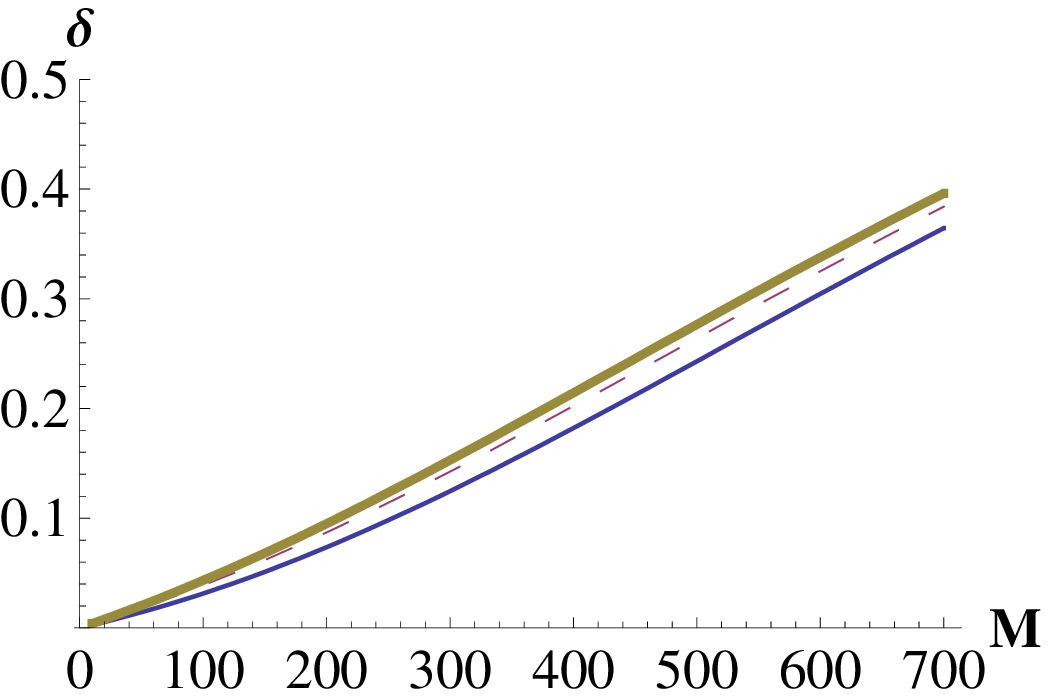}}}
    \caption{{\it Left panel}: $\delta$ as a function of the mass $M$ (in MeV) for three different chemical potentials and magnetic fields, namely $\mu=400$ MeV (thick solid line), $\mu=500$ MeV (thin solid line), and $\mu=600$ MeV
     (dashed line). In all three cases, the magnetic fields have been set to $\sqrt{2qB}=\mu$, which is the lower possible magnetic field in order to populate only the lowest Landau level. {\it Right panel}: $\delta$ as a function of $M$ for a fixed $\mu=500$ MeV, for three different magnetic fields, namely $\sqrt{qB}=500/\sqrt{2}$ MeV (thin solid line), 500 MeV (dashed line), and 600 MeV (thick solid line).
    }
    \end{center}
\end{figure}
Now we discuss the exchange amplitude of the two interacting particles. In this case $E_1=E_4$ and $E_2=E_3$. Similar relations hold also for the momenta along the $z$ and $y$ axis. The integral that corresponds to Eq.~(\ref{integ}) now becomes,
\beq \int \frac{dq_x}{2\pi}\frac{1}{q^2-M^2}\exp[-iq_x(x-x')]=-\frac{\exp[-\zeta|x-x'|]}{2\zeta}, \eeq
where $\zeta=(q_y^2+q_z^2+M^2-q_0^2)^{1/2}$. The Landau interacting function is
\beq \hat{f}(p_y,p_z;p_{y'},p_{z'})=\frac{g^2qB}{2\pi} \int dxdx'\frac{e^{-\zeta|x-x'|}}{\zeta} \exp[K(x,x',p_y,p_{y'})]  \frac{m^2}{EE'}, \label{fex}\eeq where
\beq K(x,x',p_y,p_{y'})=-\frac{qB}{2}[(x-\frac{p_y}{qB})^2+(x-\frac{p_{y'}}{qB})^2+(x'-\frac{p_{y}}{qB})^2+(x'-\frac{p_{y'}}{qB})^2]. \label{kex}\eeq As in the previous case, we are looking for solutions of the form of Eq.~(\ref{sol}) for the Boltzmann  Eq.~(\ref{eq}). The dispersion relation reads \beq (\omega - {\bf v.k}) \hat{\nu}(n)={\bf v.k}\frac{E_F}{p_F} \left ( \int \frac{dp_{y'}}{4\pi^2}[ \hat{f}(p_y,p_z;p_{y'},p_F n)\hat{\nu}(n)+ \hat{f}(p_y,p_z;p_{y'},-p_F n)\hat{\nu}(-n)] \right ). \label{disex}\eeq By using Eqs.~(\ref{fex}, \ref{kex}, \ref{disex}), we can perform the integrations over $dp_{y'}dxdx'$. Eq.~(\ref{disex}) reads
\beq (\omega \mp v_F k)\hat{\nu} (n)= \pm v_F k (f_1 \hat{\nu} (n) + f_2\hat{\nu} (-n)), \eeq
where \beq f_{1,2}= \frac{g^2}{4\sqrt{2} \pi^{3/2}}\frac{m^2}{E_F p_F}\sqrt{qB}
\left ( \frac{e^{\zeta^2_{1,2}/(2qB)}}{\zeta_{1,2}} \text{Erfc} \left [ \frac{\zeta_{1,2}}{\sqrt{2qB}} \right ] \right ). \eeq As we have mentioned $\zeta = (q_y^2+q_z^2+M^2-q_0^2)^{1/2}$. $q_0=0$ (since both the colliding particles are on the Fermi surface). It is easy to show that the integral of Eq.~(\ref{fex}) does not depend on $p_y$, and therefore we can set it to zero. On the other hand, the integral over $dp_{y'}$ is dominated by the values of $p_{y'}$ close to $p_y$. For this reason, and in order to obtain an analytic result, we have neglected the $p_{y'}$ dependence of $\zeta$. In the case where the colliding particles are moving in opposite directions,
 $q_z=2p_F>>q_y$. Therefore this approximation is justified. If the colliding particles are moving
  towards the same direction, $q_z=0$, and this approximation is justified only if $q_y<<M$. According to this, $\zeta_1=M$ (for same direction collision) and $\zeta_2=(4p_F^2+M^2)^{1/2}$ (for opposite direction collision). The dispersion relation is given again by Eq.~(\ref{omega}), where in this case $x=f_1$ and $\delta=f_2/f_1$. However unlike the previous case where $\delta>>1$, here $\delta<1$ as it can be seen in Fig.~1. Due to this, as long as $\delta<1$, the zero sound waves (provided there is no direct amplitude) are undamped. Again the zero sound velocity should be $v_0<1$. This condition leads to a constraint in $g$ that is depicted in Fig.~2.

  \begin{figure}[!tbp]
  \begin{center}

      \subfigure{\resizebox{!}{4.8cm}{\includegraphics{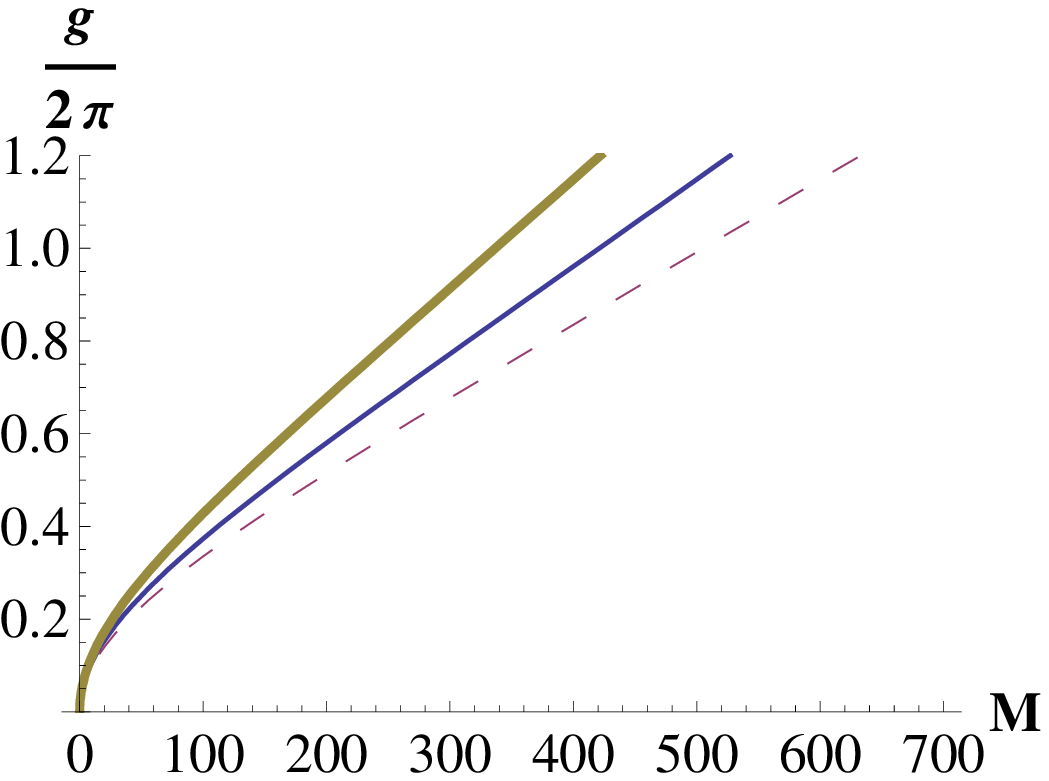}}} \quad
      \subfigure{\resizebox{!}{4.8cm}{\includegraphics{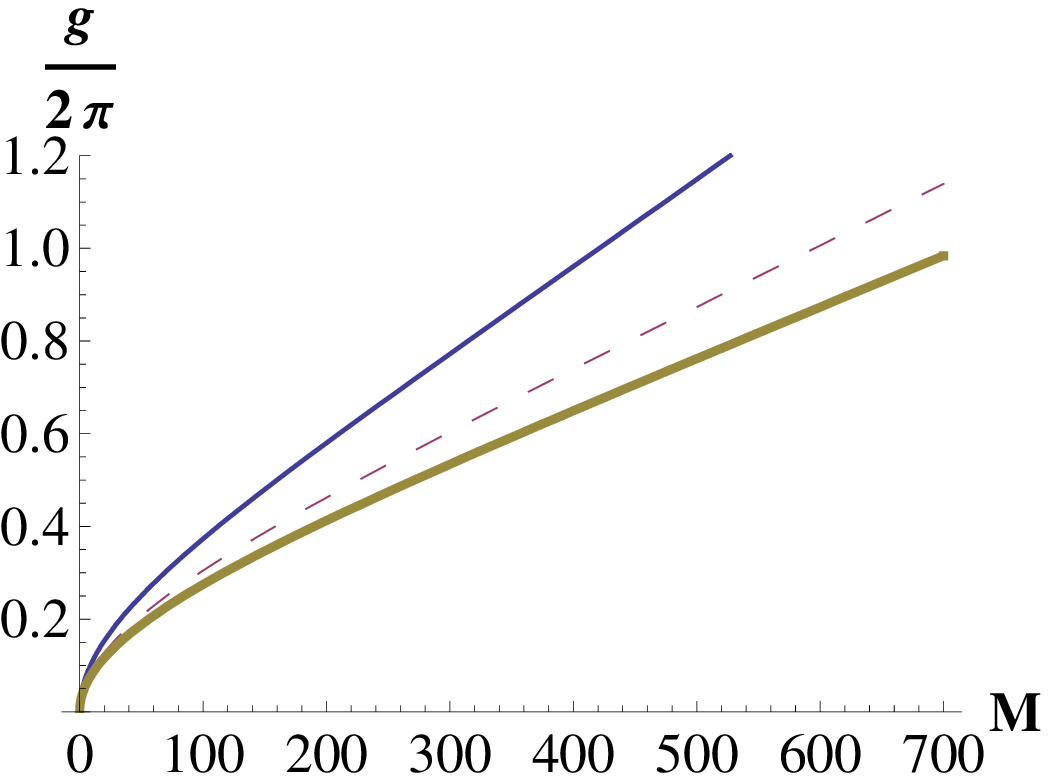}}}
    \caption{{\it Left panel}: An upper bound on $g/(2\pi)$ as a function of the mass $M$ (in order $v_0<1$) for the parameters chosen in the left panel of Fig.~1. {\it Right panel}: An upper bound on $g/(2\pi)$ as a function of the mass $M$ (in order $v_0<1$) for the chosen parameters of the right panel of Fig.~1.
    }
    \end{center}
\end{figure}

  In principle, instead of deriving the dispersion relations separately for the direct and exchange amplitudes, we should have added the two contributions $\hat{f}=\hat{f}_{Dir}+\hat{f}_{Exc}$. However, we investigated separately the two cases for two reasons. Firstly, there are interactions like QCD that do not have contributions from the direct amplitude due to the overall gauge structure. For example, in QCD the direct amplitude is zero because it will be proportional to the trace of a single Gell-Mann matrix which is zero. On the other hand, in case where both amplitudes are nonzero, by inspection of the $x$ below Eq.~(\ref{dispersion2})~(let's call it $x_{Dir}$) and $f_1$, we can conclude that for most of the cases $x_{Dir}>f_1$. This is depicted in Fig.~3. In fact we can see that for low $M$ the ratio becomes large and therefore the exchange amplitude is small compared to the direct one. In this case Eqs.~(42,~43) still hold. For larger $M$, $f_1 \sim x_{Dir}$. However, in any case $\delta_{Dir}>>\delta_{Exc}$. Therefore Eq.~(42) should  be rewritten in the general case as
  \beq x<\frac{1}{\delta-a}, \eeq where $a=1+f_1/x_{Dir}.$ In Fig.~3 we see that $1<a<2$. As before, for $\delta>>a$, or equivalently for $m<<E_F$, the constraint reduces back to Eq.~(43). It is safe to say that when both present, the direct amplitude dominates over the exchange one.
  \begin{figure}[!tbp]
\begin{center}
\includegraphics[width=0.7\linewidth]{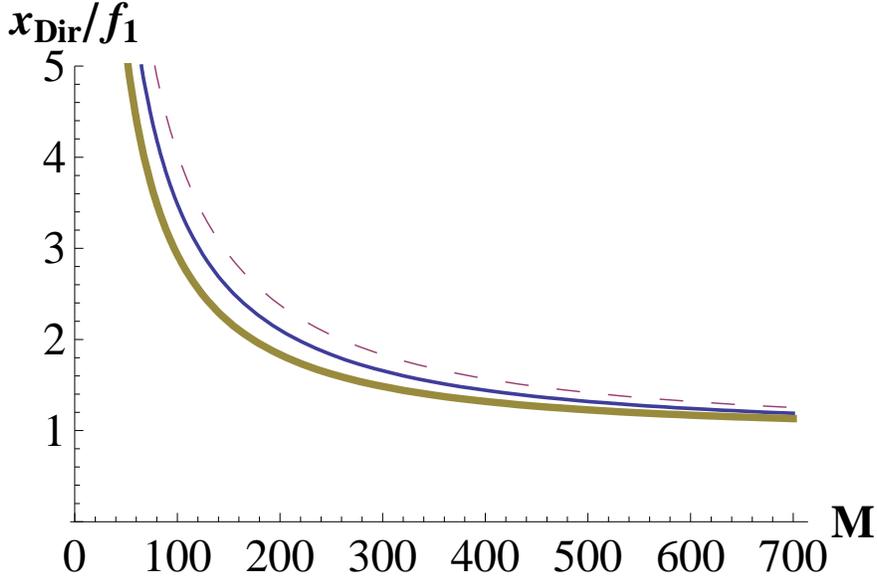}
\caption{The ratio $x_{Dir}/f_1$ as a function of $M$ for the set of parameters chosen in the left panel of Fig.~1.}
\end{center}
\end{figure}


As it can be seen in Fig.~3, for large $M$, $x_{Dir}/f_1 \rightarrow 1$. This is expected, because large $M$ means small range for the interaction, and therefore direct and exchange amplitudes become similar when momenta are aligned, since the interaction tends to be very localized (due to large $M$). The same is not true if the interacting particles have opposite momenta. One can easily check that direct and exchange amplitudes behave differently in this case even at large $M$, because of the different Dirac structure. At the chiral limit $m=0$, the exchange amplitude is zero, because the vector boson cannot flip the chirality, whereas in the direct amplitude there is no need for flipping the chirality. In fact at the chiral limit all amplitudes except the direct one with opposite momenta for the interacting particles are zero. The reason why for example the direct amplitude with aligned momenta $p_z=p'_z$ is zero at the chiral limit whereas for $p_z=-p'_z$ is not, has to do with the simple fact that particles move with the speed of light and due to the localized interacting force, they can never interact if $p_z=p'_z$ unless they are on top of each other. If $p_z=-p'_z$, they will always meet no matter how far they start from each other, provided their transverse distance is within $1/M$. We should also mention that $\delta$ in Fig.~1 saturates to 1 for larger $M$ than the ones depicted in the figure.
\section{Discussion}

In this paper we studied if the direct Urca process can occur inside a neutron star with dense quark matter under extremely high magnetic fields. When $qB \sim \mu^2$, only the lowest Landau level is populated. As we showed, the lack of transverse momentum makes impossible to satisfy simultaneously energy and momentum, and apart from very specific cases that we pointed out, direct Urca cannot take place in general. In search for finding ways to save the Urca process, we investigated sound modes and in particular zero sound. We showed that in principle if such modes are undamped, they can help facilitating a variant of the Urca process, that in some sense is a type of Modified Urca. Furthermore, by using an effective model where fermions can interact via the exchange of a vector boson, we found under what constraints, we can have undamped zero sound. We derived the dispersion relations and we calculated the zero sound velocity. We found that unlike the ``usual'' 3-dimensional case (with absent magnetic field) where $v_0>v_F$ in order to have undamped waves, in the case studied here, $v_0$ can be even smaller than $v_F$. This is due to the fact that there is no continuum in this case, since the Fermi sphere has collapsed to an 1-dimensional column, and therefore modes with velocity smaller than $v_F$ cannot dissolve back to a particle and a hole.
 \acknowledgments
 I would like to thank Chris Pethick for countless discussions, and for reading the manuscript, as well as Kenji Fukushima for useful comments.
This work is supported by the Marie Curie
Fellowship under contract MEIF-CT-2006-039211.

\end{document}